\documentclass[12pt]{article}
\usepackage{amssymb}

\usepackage{pifont}
\usepackage{mathrsfs}
\usepackage{amsfonts}
\usepackage{color}
\usepackage{footmisc}
\usepackage[numbers,sort&compress]{natbib}
\usepackage[bookmarksopen=true, bookmarksopenlevel=\maxdimen,
bookmarksnumbered=true,colorlinks,linkcolor=blue,
citecolor=blue,urlcolor=blue]{hyperref}

\newcommand{\omits}[1]{}

\begin{document}

\begin{center}
{\bf \LARGE $R+S^2$ theories of gravity without\\ \bigskip
big-bang singularity}

\bigskip

\bigskip

{\Large Jia-An Lu$^{a}$\footnote{Email: ljagdgz@163.com}}

\bigskip

$^a$ School of Physics and Engineering, Sun Yat-sen
University,\\ Guangzhou 510275, China

\begin{abstract}
The $R+S^2$ theories of gravity, where $S^2$ denotes the quadratic torsion terms,
are analyzed under three cases.
In the first two cases, the matter fields are described by two different
spin fluids which are not homogeneous and isotropic.
In the third case, a homogeneous and isotropic torsion field is used. It is found that
under all the three cases, the $R+S^2$ theories may avert the big-bang
singularity of the Robertson--Walker universe, with three corresponding constraints
on the parameters.
\end{abstract}
\end{center}



\section{Introduction}

The spacetime torsion is introduced in the gauge theory of gravity
\cite{Hehl76,Guo76,Stelle,Grignani,PoI,Guo07,SR-Gravity,Lu13,Lu14,Lu14-b} to
realize the local Poincar\'e, de Sitter (dS) or Anti-de Sitter (AdS) symmetry.
It is shown that the torsion effect in the Einstein--Cartan (EC) theory,
which is the simplest model of the gauge theory of gravity, may avert the
initial singularity of the homogeneous but anisotropic universe \cite{Kuchowicz1},
where the matter fields are described by a spin fluid \cite{Kuchowicz2}.
It is also shown that EC theory may avert the big-bang singularity of the Robertson--Walker
(RW) universe \cite{Kuchowicz3,Poplawski}, where the matter fields are represented
by another spin fluid \cite{Hehl74,Obukhov}. In these works
\cite{Kuchowicz3,Poplawski}, an averaging procedure \cite{Hehl74}
is used such that the torsion field is equal to zero but the torsion effect does not vanish.
However, this averaging procedure is ambiguous for the classical gravitation
theory \cite{Lu14-b}. Without the use of this averaging procedure, the torsion fields are
actually anisotropic in these works. On the other hand, it is shown that if the
torsion field is homogeneous and isotropic, the EC theory cannot avert the big-bang
singularity of the RW universe \cite{Tunyak}.

Recently, the singularity problem of the RW universe is analyzed in
an extension of the EC theory: the $R+\beta S^{abc}S_{abc}$ models of gravity,
where $\beta$ is a parameter, $S_{abc}$ denotes torsion.
When the matter fields are described by the first-mentioned spin fluid \cite{Kuchowicz2}
(spin fluid A), it is found that only the model with $\beta=1/2$ may avoid the initial
singularity \cite{Lu14-b}. When the second-mentioned spin fluid \cite{Hehl74,Obukhov}
(spin fluid B) is used, it is found that the models may avoid the singularity while
$\beta$ satisfies some constraint \cite{Vognolo,Lu14-b}.

As a matter of fact, the $R+\beta S^{abc}S_{abc}$ models of gravity belong to the
general $R+S^2$ theories of gravity
\cite{S2-11,S2-12}, where $S^2$ stands for all possible quadratic torsion terms.
The $R+S^2$ theories of gravity are complete in the sense that the Lagrangians contain
both curvature and torsion. It is an interesting investigation, which will be done
in this paper, to find which $R+S^2$ models of gravity may solve
the singularity problem of the RW universe under the aforementioned three
cases: (a) the matter fields are represented by the spin fluid A \cite{Kuchowicz2};
(b) the matter fields are represented by the spin fluid B \cite{Hehl74,Obukhov};
(c) the torsion field is homogeneous and isotropic.

The paper is organized as follows. In section 2, the $R+S^2$ theories of
gravity are briefly introduced. In sections 3--5, it is found that the torsion effect
in the $R+S^2$ theories may avoid the initial singularity of the RW universe under
the above-mentioned three cases, with three corresponding constraints
on the parameters of the $R+S^2$ theories. Finally some remarks are given
in the last section.

\section{\texorpdfstring{$R+S^2$}{R+S2} theories of gravity}

We will consider the $R+S^2$ theories of gravity as follows \cite{S2-11,S2-12}:
\begin{equation}\label{LG}
\mathscr{L}_G=R-2\Lambda+a_1S_{cab}S^{cab}+a_2S_{cab}S^{acb}+a_3S_aS^a,
\end{equation}
where the notation of Ref. \cite{Lu14} is used, $\mathscr L_G$ is the gravitational
Lagrangian, $R$ is the scalar curvature, $\Lambda$ is the positive cosmological constant,
$S_{cab}=g_{cd}S^d{}_{ab}$, $S^d{}_{ab}$ is the torsion tensor, $g_{cd}$ is the metric
tensor, $S_a=S^c{}_{ac}$, $a,b,c,$ etc., are abstract indices \cite{Wald,Liang},
and $a_1,a_2,a_3$ are three dimensionless parameters. The Lagrangian (\ref{LG}) is
gauge invariant because each of the metric, torsion and
curvature can be expressed in a gauge-invariant way \cite{Guo76,Stelle,Grignani,Lu13,Lu14,Lu14-b}.
Moreover, the Lagrangian is complete in the sense that it contains all components
of the gravitational field strength ${\cal F}_{ab}$, i.e., it contains both curvature and
torsion. In fact, when the gravitational gauge group is chosen to be the dS group,
\begin{equation}
{\cal F}_{ab}=\left(
\begin{array}{cc}
R^{\alpha}{}_{\beta ab}-l^{-2}e^{\alpha}{}_{a}\wedge e_{\beta b}
&l^{-1}S^{\alpha}{}_{ab}\\
-l^{-1}S_{\beta ab}&0
\end{array}
\right)
\end{equation}
in a special gauge \cite{Lu13,Lu14}, where $R^{\alpha}{}_{\beta ab}$
is the curvature 2-form with respect to an orthonormal frame field
$\{e_\alpha{}^a\}$, $S^{\alpha}{}_{ab}$ is the torsion 2-form with respect
to $\{e_\alpha{}^a\}$, $\{e^{\alpha}{}_{a}\}$ is the dual of
$\{e_\alpha{}^a\}$, $e_{\beta b}=\eta_{\alpha\beta}e^\alpha{}_b$,
$S_{\beta ab}=\eta_{\alpha\beta}S^\alpha{}_{ab}$, $\alpha,\beta=0,1,2,3$,
and $l>0$ is related to the cosmological constant by $\Lambda=3/l^2$. Also,
the Lagrangian (\ref{LG}) is simple in the sense
that it is comprised of the simplest curvature and torsion scalars.

The gravitational field equations corresponding to the Eq. (\ref{LG})
are as follows:
\begin{eqnarray}\label{1stEq}
R_{ba}-\frac1 2Rg_{ab}+\Lambda g_{ab}-\frac1 2
(a_1S_{cde}S^{cde}+a_2S_{cde}S^{dce}+a_3S_cS^c)g_{ab}\nonumber\\
+2a_1S^{cd}{}_aS_{cdb}-a_2S^{cd}{}_aS_{bcd}+a_2S^{cd}{}_aS_{dcb}
+a_3S_aS_b-a_3S_{bac}S^c\nonumber\\
-2\nabla_c(a_1S_{ab}{}^{c}+\frac1 2a_2S_{ba}{}^c-\frac1 2a_2S^c{}_{ab}
+\frac1 2a_3\delta^c{}_aS_b-\frac1 2a_3g_{ab}S^c)\nonumber\\
-T_{bcd}(a_1S_a{}^{cd}-a_2S^{[cd]}{}_a+a_3S^{[c}\delta^{d]}{}_a)
=\frac{1}{2\kappa}\Sigma_{ab},
\end{eqnarray}
\begin{equation}\label{2ndEq}
 T^{a}{}_{bc}+(2a_2-4a_1)S_{[bc]}{}^{a}+2a_2S^a{}_{bc}
 -2a_3\delta^a{}_{[b}S_{c]}=-\frac{1}{\kappa}\tau_{bc}{}^{a},
\end{equation}
where $R_{ba}$ is the Ricci tensor, $\nabla_c$ is the metric-compatible
derivative operator, $T^a{}_{bc}=S^a{}_{bc}+2\delta^a{}_{[b}S_{c]}$,
$\kappa$ is the gravitational coupling constant, $\Sigma_{ab}$ is the
canonical energy-momentum tensor, and $\tau_{bc}{}^a$ is the spin tensor.
Note that the cases with $2a_1+a_2+3a_3=2$, $2a_1+a_2+1=0$, or $4a_1-4a_2-1=0$
are not so general for the reason that they can only describe the matter fields with
$\tau_b\equiv\tau_{bc}{}^c=0$, $g_{a[b}\tau_{c]}-\tau_{a[bc]}+\tau_{bca}=0$,
or $\tau_{[abc]}=0$. For the cases with $2a_1+a_2+3a_3\neq2$, $2a_1+a_2+1\neq0$,
and $4a_1-4a_2-1\neq0$ , the solution of Eq. (\ref{2ndEq}) is
\begin{eqnarray}\label{tor-spin}
S_{abc}=(2a_1+a_2+1)^{-1}(4a_1-4a_2-1)^{-1}[(2-2a_3)(4a_2-4a_1+1)g_{a[b}S_{c]}\nonumber\\
-(2a_2-4a_1)\kappa^{-1}\tau_{a[bc]}+(3a_2-2a_1+1)\kappa^{-1}\tau_{bca}],
\end{eqnarray}
where $S_c=\tau_c/[\kappa(2-2a_1-a_2-3a_3)]$.
The symmetric energy-momentum tensor $T_{ab}$ is related to the canonical energy-momentum tensor
$\Sigma_{ab}$ and the spin tensor $\tau_{bc}{}^a$ by
\begin{eqnarray}\label{T-Sigma}
T_{ab}&=&\Sigma_{ab}+(\nabla_c+S_c)(\tau_{ab}{}^c-\tau_a{}^c{}_b+\tau^c{}_{ba})\nonumber\\
&=&\Sigma_{(ab)}+2(\nabla_c+S_c)\tau^c{}_{(ab)}.
\end{eqnarray}
Hence, if the spin tensor is equal to zero, then $S^c{}_{ab}=0$,
$\Sigma_{ab}=T_{ab}$, and Eq. (\ref{1stEq}) reduces to the Einstein field
equation with a cosmological constant
when $1/2\kappa=8\pi$.
Generally, Eq. (\ref{1stEq})
can be expressed as an Einstein-like equation:
\begin{equation}\label{Elike}
\mathring R_{ab}-\frac1 2\mathring Rg_{ab}+\Lambda g_{ab}
=\frac1{2\kappa}(T_{\rm eff})_{ab},
\end{equation}
where $\mathring R=g^{ab}\mathring R_{ab}$, $\mathring R_{ab}$
is the torsion-free Ricci tensor, and $(T_{\rm eff})_{ab}$ is the effective
energy-momentum tensor which satisfies
\begin{eqnarray}\label{Teff}
\frac1{2\kappa}(T_{\rm eff})_{ab}=\frac1{2\kappa}T_{ab}
-\frac1\kappa(\frac12S_{(a}{}^{cd}\tau_{|cd|b)}+K^d{}_{(a|c|}\tau^c{}_{b)d})
+\frac1 4[(\frac1 2+2a_1)S_{cde}S^{cde}\nonumber\\
+(1+2a_2)S_{cde}S^{dce}+2(a_3-1)S_cS^c]g_{ab}
+\frac12(1+2a_1+a_2)S^{cd}{}_{(a}S_{b)cd}\nonumber\\
+(a_3-1)S_{(ab)}{}^cS_c-\frac14(1+2a_2)S_{acd}S_b{}^{cd}
+(a_2-2a_1)S_{cda}S^{[cd]}{}_b,
\end{eqnarray}
where $K^d{}_{bc}$ is the contorsion tensor. It is remarkable that
all the derivative terms of the torsion tensor are absorbed into
$T_{ab}$, and therefore $(T_{\rm eff})_{ab}$
is different from $T_{ab}$ only by some quadratic torsion terms.

\section{Spin fluid A}
Let us assume that the matter fields in the universe can be described by
the spin fluid A \cite{Kuchowicz2} with the symmetric energy-momentum tensor and
spin tensor being
\begin{equation}\label{T}
T_{ab}=\rho U_aU_b+p(g_{ab}+U_aU_b),
\end{equation}
\begin{equation}\label{tau}
\tau_{bc}{}^a=\tau_{bc}U^a,
\end{equation}
where $\rho$ is the rest energy density, $p$ is the hydrostatic pressure,
$U^a$ is the four-velocity of the fluid particles, and $\tau_{bc}$ is the spin
density 2-form which satisfies $\tau_{bc}U^c=0$. It is an extension of
the special relativistic Weyssenhoff fluid \cite{Weyssenhoff}. Substitution of
Eq. (\ref{tau}) into Eq. (\ref{Teff}) yields
\begin{eqnarray}\label{Teff2}
\frac1{2\kappa}(T_{\rm eff})_{ab}=\frac1{2\kappa}T_{ab}
+\frac1{\kappa^2}(2a_1+a_2+1)^{-2}(4a_1-4a_2-1)^{-2}\nonumber\\
\times[A\tau_a{}^c\tau_{bc}+Bs^2U_aU_b+Cs^2g_{ab}],
\end{eqnarray}
where $s^2=\tau_{bc}\tau^{bc}/2$ is the spin density square, and
\begin{equation}
A=\frac12(2a_1-3a_2-1)(2a_1+a_2+1)(4a_1-4a_2-1),
\end{equation}
\begin{equation}
B=-2(2a_1-a_2)^3+\frac12(2a_1-3a_2-1)^2(4a_1+1),
\end{equation}
\begin{eqnarray}\label{C}
C=\frac12[-2(2a_1-a_2)^3-\frac12(2a_1-3a_2-1)^2(4a_1+1)\nonumber\\
-(2a_1-3a_2-1)(4a_1-2a_2)(2a_2+1)].
\end{eqnarray}
Furthermore, the metric field of the universe
is supposed to be an RW metric with the line element
\begin{equation}\label{RW}
ds^2=-dt^2+a^2(t)[dr^2+r^2(d\theta^2+\sin^2\theta d\varphi^2)],
\end{equation}
where $t$ is the cosmic time with $(\partial/\partial t)^a=U^a$.
According to Eq. (\ref{RW}), the left-hand side of
the Einstein-like equation (\ref{Elike}) is diagonal in the coordinate system
$\{t,r,\theta,\varphi\}$, then $(T_{\rm eff})_{ab}$
should be diagonal too, and so the term $A\tau_{ac}\tau_b{}^c$ in Eq.
(\ref{Teff2}) should be diagonal, which implies that $A=0$ or $\tau_{bc}=0$.
In order to solve the singularity problem of the RW universe, the Einstein-like
equation (\ref{Elike}) should be different from the Einstein field equation.
Consequently, $\tau_{bc}\neq0$, and so $A=0$, which implies
\begin{equation}\label{a1}
2a_1-3a_2-1=0.
\end{equation}
Substitution of Eq. (\ref{a1}) into Eq. (\ref{Teff2}) leads to
\begin{equation}\label{Teff3}
(T_{\rm eff})_{ab}=(\rho+\rho_S)U_aU_b+(p+p_S)(g_{ab}+U_aU_b),
\end{equation}
where
\begin{equation}\label{state}
\rho_S=p_S=-8\pi s^2/(2a_2+1).
\end{equation}
It is seen that the torsion effect is equivalent to an
ideal fluid with the rest energy density equal to the hydrostatic pressure.

Note that the torsion field given by Eqs. (\ref{tor-spin}) and (\ref{tau}) is
not homogeneous and isotropic, since the spin tensor (\ref{tau}) is not homogeneous
and isotropic. It is unnatural to assume that the spin tensor is homogeneous, in which case
there is a special direction given by $\tau_{bc}$ for the whole universe.
Also, the spin tensor cannot be isotropic because there is a special direction given by
$\tau_{bc}$ at each point of the universe. The condition (\ref{a1}) essentially tells
in which models an anisotropic spin fluid can exist in the RW universe.

Substituting the RW line element (\ref{RW}) and Eq. (\ref{Teff3}) into the
Einstein-like equation (\ref{Elike}) results in the dynamical equations of the universe:
\begin{equation}\label{Flike}
(\dot{a}/a)^2=(8\pi/3)(\rho+\rho_S+\rho_\Lambda),
\end{equation}
\begin{equation}\label{addot}
\ddot{a}/a=(-4\pi/3)(\rho+3p+4\rho_S-2\rho_\Lambda),
\end{equation}
where $\rho_\Lambda=\Lambda/8\pi$. The Friedmann-like equation (\ref{Flike})
together with Eq. (\ref{state}) implies that the spin density scalar
$s$ is a function of the cosmic time. Moreover, a conservation law can be
deduced from Eqs. (\ref{Flike}) and (\ref{addot}):
\begin{equation}\label{conservation}
\dot{\rho}+\dot{\rho}_S+(3\dot a/a)(\rho+p+2\rho_S)=0.
\end{equation}
When $\dot a\neq0$, Eq. (\ref{addot}) can be replaced by the conservation
law (\ref{conservation}). In the early universe, $p=\rho/3$, then Eq. (\ref{conservation})
has the following solution: $\rho\propto a^{-4}$, and
\begin{equation}\label{rhoS2}
\rho_S\propto a^{-6}.
\end{equation}
Also, $\rho_\Lambda$ can be neglected in the early universe, then $\dot a=0$
can be realized at the moment when $\rho+\rho_S=0$. Set $t=0$ at this moment,
and define $s_b=s(0)$, $\rho_b=\rho(0)$, and $\rho_{Sb}=\rho_S(0)$.
Hence $\rho_b+\rho_{Sb}=0$, which requires $\rho_{Sb}<0$, i.e., $s_b\neq0$, and
\begin{equation}\label{a2}
a_2>-\frac12.
\end{equation}
In conclusion, the torsion effect may replace the big-bang singularity with a
big bounce, to ensure which the parameters should satisfy Eqs. (\ref{a1})
and (\ref{a2}). When the universe expands, the torsion effect would rapidly
decay due to Eq. (\ref{rhoS2}).

Suppose that $s=Hn$, where $H$ is a dimensional constant, and $n=n(t)$ is
the total number density of all fundamental fermions and massive bosons.
For example, one may let $H=\hbar/2$ for
an electron fluid with aligned spins. Here it is assumed that only fermions
and massive bosons are coupled to torsion, in order to preserve the gauge invariance
with respect to the massless bosons \cite{Hehl76}. In the early universe, the energy density
and number density satisfy \cite{Kolb} $\rho=C_\rho T^4$ and $n=C_n T^3$, respectively,
where $T=T(t)$ is the temperature,
$C_\rho$ and $C_n$ are constants which only depend on the number of spin states of
each species of particles. With the help of Eq. (\ref{state}) and the above assumptions,
it yields that $\rho_b=-\rho_{Sb}=8\pi H^2n_b^2/(2a_2+1)$, where $n_b=n(0)$, and so
\begin{equation}
T_b^2=(2a_2+1)C_\rho/8\pi H^2C_n^2,
\end{equation}
where $T_b=T(0)$. By adjusting the parameter $a_2$, there could be $T_b\ll T_P$,
where $T_P$ is the Planck temperature, such that the quantum gravity effects
could be neglected.

\section{Spin fluid B}

Apart from the spin fluid A, there is another
extension of the special relativistic Weyssenhoff fluid:
the spin fluid B \cite{Hehl74,Obukhov},
which satisfies Eq. (\ref{tau}) and
\begin{equation}\label{Sigma}
\Sigma_{ab}=W_aU_b+p(g_{ab}+U_aU_b),
\end{equation}
where $W_a$ is the momentum density. Utilizing the spin conservation law
$(\nabla_c+S_c)\tau_{ab}{}^c=-\Sigma_{[ab]}$ and Eq. (\ref{tau}) gives
\begin{equation}\label{Sigma2}
\Sigma_{ab}=\rho U_aU_b+p(g_{ab}+U_aU_b)+2U_bU^cU^d\mathring\nabla_c\tau_{ad},
\end{equation}
where $\rho=-W_aU^a$ is the rest energy density. Substitution of Eqs. (\ref{Sigma2})
and (\ref{tor-spin}) into Eq. (\ref{T-Sigma}) leads to
\begin{eqnarray}\label{T2}
T_{ab}=\rho U_aU_b+p(g_{ab}+U_aU_b)+2\mathring\nabla_c\tau^c{}_{(ab)}
+2U^cU^dU_{(b}\mathring\nabla_{|c|}\tau_{a)d}\nonumber\\
+\frac{3a_2-2a_1+1}{(2a_1+a_2+1)(4a_1-4a_2-1)}
\frac1\kappa(2s^2U_aU_b+\tau_a{}^c\tau_{bc}).
\end{eqnarray}
Substituting the above equation into Eq. (\ref{Teff2}) yields
\begin{eqnarray}\label{Teff4}
(T_{\rm eff})_{ab}=(\rho+\rho_S)U_aU_b+(p+p_S)(g_{ab}+U_aU_b)\nonumber\\
+2\mathring\nabla_c\tau^c{}_{(ab)}+2U^cU^dU_{(b}\mathring\nabla_{|c|}\tau_{a)d},
\end{eqnarray}
where
\begin{eqnarray}\label{rhoS}
\rho_S=p_S=(2/\kappa)(2a_1+a_2+1)^{-2}(4a_1-4a_2-1)^{-2}Cs^2.
\end{eqnarray}
For the RW universe, the last term of Eq. (\ref{Teff4}) is equal to zero,
and the nonzero components of the term $2\mathring\nabla_c\tau^c{}_{(ab)}$
in Eq. (\ref{Teff4}) are \cite{Lu14-b}:
\begin{equation}\label{curl}
2\mathring\nabla_c\tau^c{}_{(0i)}=\hat\nabla^j\hat\tau_{ij}
=\hat\varepsilon_{ijk}\hat\nabla^j(^*\hat\tau)^k\equiv[{\rm curl}(^*\hat\tau)]_i,
\end{equation}
where $\hat\nabla_i$ is the $\hat g_{ij}$-compatible
torsion-free derivative operator, $\hat g_{ij}$ is the confinement of $g_{ab}$
on the cosmic space $\Sigma_t$, $\hat\tau_{ij}$ is the confinement of $\tau_{bc}$
on $\Sigma_t$, $\hat\varepsilon_{ijk}$ is the Levi-Civita symbol,
$(^*\hat\tau)_k=\hat\tau^{ij}\hat\varepsilon_{ijk}/2$ is the spin density 1-form, and,
$i,j,k=1,2,3$ are spatial indices and raised by the inverse of $\hat g_{ij}$.
Since the off-diagonal
term of $(T_{\rm eff})_{ab}$ should be equal to zero, the spin density vector
$(^*\hat\tau)^i$ should satisfy
\begin{equation}\label{startau}
{\rm curl}(^*\hat\tau)=0.
\end{equation}
As mentioned before, the spin tensor is not homogeneous and isotropic.
The condition (\ref{startau}) essentially tells which kind of inhomogeneous
and anisotropic spin fluid can exist in the RW universe.
Actually, the term $2\mathring\nabla_c\tau^c{}_{(ab)}$ in Eq. (\ref{Teff4})
is usually discarded by an averaging argument \cite{Hehl74,Poplawski,Kuchowicz3,Vognolo}:
if the microscopic spin orientations are random, then the spin tensor and its derivative
vanish after macroscopic averaging. However, the averaging procedure is ambiguous,
since the gravitational field equations cannot be averaged in any classical theory of
gravity \cite{Lu14-b}.

It can be seen from Eqs. (\ref{Teff4}), (\ref{rhoS}) and (\ref{startau})
that the torsion effect is again equivalent to an ideal fluid with the
state equation $\rho_S=p_S$. Similar to the previous section, it can be shown that
the torsion effect in the present case may also avert the big-bang singularity,
and the parameters should be constrained by $C<0$, where $C$ is given by Eq. (\ref{C}).
For example, if $a_2=0$, then
\begin{equation}
C=(4a_1-1)(2a_1+1)(-6a_1+1)/4.
\end{equation}
To ensure $C<0$, there should be $-1/2<a_1<1/6$ or $a_1>1/4$. For another example,
if Eq. (\ref{a1}) is satisfied, then
\begin{equation}
C=-(2a_2+1)^3,
\end{equation}
and Eq. (\ref{rhoS}) is the same as Eq. (\ref{state}). Generally, define
\begin{equation}
\overline C=-4C(2a_1+a_2+1)^{-2}(4a_1-4a_2-1)^{-2},
\end{equation}
then Eq. (\ref{rhoS}) can be expressed as
\begin{equation}
\rho_S=p_S=-8\pi\overline C s^2.
\end{equation}
As before, suppose that $s=Hn$. Making use of the relations $\rho=C_\rho T^4$
and $n=C_nT^3$, it follows that the temperature at the big bounce with $\dot a=0$
satisfies
\begin{equation}
T_b^2=C_\rho/8\pi H^2C_n^2\overline C.
\end{equation}
By adjusting the parameters $a_1$ and $a_2$, there could be $T_b\ll T_P$, such that
the quantum gravity effects could be neglected. In comparison, this cannot yet be realized
in the EC theory with $a_1=a_2=0$ \cite{Poplawski2}.

\section{Homogeneous and isotropic torsion}

As mentioned before, the spin fluids used in the previous sections are not
homogeneous and isotropic. It remains a problem how to derive a spin fluid from the
basic physical laws such that it can be homogeneous and isotropic. However,
we may start from the geometrical parts, i.e., the metric and torsion fields.
It can be shown that \cite{Tunyak,Tsamparlis}
the homogeneous and isotropic torsion can be expressed as
\begin{equation}\label{Hitor}
S^c{}_{ab}=2b(t)\delta^c{}_{[a}U_{b]},
\end{equation}
where $b(t)$ is a function to be determined. Substituting the above equation into
Eq. (\ref{2ndEq}) leads to
\begin{equation}\label{tau2}
\tau_{bc}{}^a=\kappa(2-2a_1-a_2-3a_3)S^a{}_{bc}.
\end{equation}
This spin tensor is different from the Weyssenhoff spin tensor (\ref{tau})
which is based on a classical description of the particles. However, it is more
reasonable to use the quantum description of the particles in the early universe.
It remains a problem whether the spin tensor (\ref{tau2}) can be based on a
quantum description of the particles.
Substitution of Eqs. (\ref{Hitor}) and (\ref{tau2}) into Eq. (\ref{Teff}) yields
\begin{equation}\label{Teff5}
(T_{\rm eff})_{ab}=T_{ab}+\rho_SU_aU_b+p_S(g_{ab}+U_aU_b),
\end{equation}
where
\begin{equation}\label{rhoS3}
\rho_S=p_S=3\kappa b^2(2-2a_1-a_2-3a_3).
\end{equation}
The RW line element (\ref{RW}) together with Eqs. (\ref{Elike}) and
(\ref{Teff5}) requires that the symmetric energy-momentum tensor $T_{ab}$ should
take the form of Eq. (\ref{T}).

According to Eqs. (\ref{Elike}), (\ref{Teff5}) and (\ref{rhoS3}),
the torsion effect is again equivalent to an ideal fluid with the state
equation $\rho_S=p_S$.
Similar to section 3, it can be shown that the torsion effect
may replace the big-bang singularity by a big bounce, and the parameters should be
constrained by
\begin{equation}
2-2a_1-a_2-3a_3<0.
\end{equation}
Because Eq. (\ref{rhoS2}) still holds for the present case, it follows from
Eq. (\ref{rhoS3}) that
\begin{equation}
b\propto a^{-3}.
\end{equation}
Observing that the number density $n\propto a^{-3}$, we may assume that $b=C_bn$,
where $C_b$ is a constant. It follows that the temperature at the big bounce satisfies
\begin{equation}
T_b^2=C_\rho/3\kappa C_b^2C_n^2(2a_1+a_2+3a_3-2).
\end{equation}
By adjusting the parameters, there could be $T_b\ll T_P$, such that the quantum gravity
effects could be neglected.

Note that for the three cases discussed in sections 3--5, it holds that $\rho_S=-C_Sn^2$,
where the constant $C_S$ depends on the parameters $a_1,\ a_2$ and $a_3$ in different
ways for the three cases. At the big bounce, $\rho+\rho_S=0$, and so $T_b^2=C_\rho/C_SC_n^2$.
Denote the moment at which $\ddot{a}=0$ by $t_1$. According to Eq. (\ref{addot}),
$\rho+2\rho_S=0$ at $t_1$, and so $T_1^2=C_\rho/2C_SC_n^2$, where $T_1=T(t_1)$.
It is seen that
\begin{equation}\label{T1Tb}
T_1^2=T_b^2/2.
\end{equation}
Although $T_b$ and $T_1$ themselves depend on the parameters $a_1,\ a_2$ and $a_3$,
the relation between them is independent of the parameters. When $T_1<T<T_b$, the
expansion of the universe is accelerated. When $T<T_1$, the expansion of the universe
is decelerated until the effect of the cosmological constant appears. When $T^2\ll T_b^2$,
it holds that $|\rho_S|\ll\rho$, and hence the torsion effect is negligible.

\section{Remarks}

It is shown that the $R+S^2$ theories of gravity may avoid the big-bang singularity
of the RW universe under three cases, with three corresponding constraints on the parameters.
It is remarkable that in the gravitational field equations, all of the
derivative terms of torsion can be absorbed into the symmetric energy-momentum tensor,
and the torsion effect is equivalent to an ideal fluid with the rest energy density
equal to the hydrostatic pressure for all the three cases. Moreover, it is found that
by adjusting the parameters, the big bounce replacing the singularity could become
a classical bounce, where the quantum gravity effects could be neglected.

\section*{Acknowledgments}
I would like to thank Profs. C.-G. Huang, Z.-B. Li and X.-W. Liu for their help
and some useful discussions. I would also like to thank Prof. Y. Obukhov for the
discussions on the spin fluid models.

\end{document}